\documentclass{sig-alternate-2013}
\newfont{\mycrnotice}{ptmr8t at 7pt}
\newfont{\myconfname}{ptmri8t at 7pt}

\permission{Permission to make digital or hard copies of all or part of this
work for personal or classroom use is granted without fee provided that
copies are not made or distributed for profit or commercial advantage and
that copies bear this notice and the full citation on the first page.
Copyrights for components of this work owned by others than ACM must be
honored. Abstracting with credit is permitted. To copy otherwise, or
republish, to post on servers or to redistribute to lists, requires prior
specific permission and/or a fee. Request permissions from
permissions@acm.org.}
\conferenceinfo{KDD'13,}{August 11--14, 2013, Chicago, Illinois, USA.}
\CopyrightYear{2013}
\crdata{978-1-4503-2174-7/13/08}

\usepackage{graphicx}
\usepackage{amsfonts,amsmath,amssymb}
\usepackage{bbm}
\usepackage{color}
\usepackage[caption=false]{subfig}
\usepackage{soul}

\newcommand\independent{\protect\mathpalette{\protect\independenT}{\perp}}
\def\independenT#1#2{\mathrel{\rlap{$#1#2$}\mkern2mu{#1#2}}}

\renewcommand\vert{\; | \;}

\newcommand\V{\text{V}}

\newcommand{\sumdot}{\text{\tiny$\bullet$}}

\begin{document}

\title{Uncertainty in Online Experiments with Dependent Data: An Evaluation of Bootstrap Methods}
\numberofauthors{2}
\author{
\alignauthor{Eytan Bakshy} \\
 \affaddr{Facebook}\\
\email{eytan@fb.com}
\alignauthor{Dean Eckles} \\
 \affaddr{Facebook}\\
\email{deaneckles@fb.com}
}

\toappear{A version of this paper appeared in \emph{Proceedings of the 19th ACM SIGKDD International Conference on Knowledge Discovery and Data Mining (KDD 2013)}.\\ \\
\sf{Since publication, we found an error in one set of simulations presented in Section 3.4 and Figure 4. This version removes those simulations and corrects related statements. All other results are unaffected.}
}

\maketitle
\begin{abstract}
Many online experiments exhibit dependence between users and items. For example, in online advertising, observations that have a user \emph{or} an ad in common are likely to be associated.   Because of this, even in experiments involving millions of subjects, the difference in mean outcomes between control and treatment conditions can have substantial variance. Previous theoretical and simulation results demonstrate that not accounting for this kind of dependence structure can result in confidence intervals that are too narrow, leading to inaccurate hypothesis tests.

We develop a framework for understanding how dependence affects uncertainty in user--item experiments and evaluate how bootstrap methods that account for differing levels of dependence  perform in practice. We use three real datasets describing user behaviors on Facebook --- user responses to ads, search results, and News Feed stories --- to generate data for synthetic experiments in which there is no effect of the treatment on average by design. We then estimate empirical Type I error rates for each bootstrap method.  Accounting for dependence within a single type of unit (i.e., within-user dependence) is often sufficient to get reasonable error rates. But when experiments have effects, as one might expect in the field, accounting for multiple units with a multiway bootstrap can be necessary to get close to the advertised Type I error rates. This work provides guidance to practitioners evaluating large-scale experiments, and highlights the importance of analysis of inferential methods for dependence structures common to online systems.
\end{abstract}

\category{G.3}{Probability and Statistics}{Statistical Computing}

\keywords{causal inference, bootstrapping, field experiments, A/A tests, A/B testing, random effects, user--item data}

\pagebreak
\section{Introduction}
Experiments conducted on the Internet frequently involve millions to tens of billions of observations. This could lead to the perception that there is little uncertainty about experimental outcomes. However, treatment effects are often very small in absolute terms, so a great number of observations can be required to distinguish them from noise. Furthermore, many Internet-scale datasets, including those generated by social media feeds, search, ads, and recommender systems, have a user--item structure such that individual observations are not independent; rather, there is substantial dependence between observations of the same units.
For example, consider an online advertising experiment in which there 1 million ad impressions, but these only include 1,000 distinct ads and 10,000 distinct users. Clearly, the effective sample size will be less than 1 million, and there can be substantial uncertainty about the difference in click-through rate (CTR) between the treatment and control.

Accounting for this dependence is important for statistical inference, including hypothesis testing and confidence interval estimation. Inferential procedures that neglect this dependence structure are expected to be anti-conservative: they will have higher Type I error rates than expected and, e.g., ``95\%'' confidence intervals will include the true value less than 95\% of the time. 

High false positive rates have substantial managerial consequences. For example, experiments using one popular experimentation platform at Facebook compare, on average, 3.7 non-control conditions.  With four comparisons of independent experiments and a nominal Type I error rate $\alpha = 0.05$, there should be a $1 - (1 - 0.05)^4 = 18.5\%$ chance that at least one condition would be significant under the null hypothesis (i.e., one in 5.4 experiments with no effects may yield at least one significant condition).  But if the true Type I error was considerably higher, say $\alpha = 0.2$, one would have a $1 - (1 - 0.2)^4 = 59\%$ chance of having falsely rejected a null hypothesis. Given that many experiments involve comparing multiple outcomes (i.e., metrics), in practice the resulting effects on decision making can be worse than this suggests: not only can there be errors in identifying effects on the primary outcome, but incorrectly rejecting the null for secondary outcomes might delay or prevent the launch of a change that is otherwise beneficial.

This paper describes sources and consequences of dependence in common applications of experimentation to Internet services. We posit a general data generating process and illustrate how experimental assignment procedures and common effects of units (e.g., users and ads) affect the true uncertainty about experimental comparisons. We then evaluate independent, one-way, and multiway bootstrap methods for computing confidence intervals using null experiments (``A/A tests'') derived from three empirical datasets from Facebook: clicks on advertisements, search results, and content in the News Feed.
To examine performance under additional deviations from the null, we conduct simulations using a realistic probit random effects model.

Our primary contribution is providing guidance about when accounting for dependence among observations are most important: while previous work has shown that neglecting all dependence structure results in massive overconfidence, less work has examined how accounting for some sources of dependence, but not others, affects inference in practice.
We conclude that analysts should  use a inferential procedure that accounts for dependence among observations of the units assigned to conditions (e.g., users), but that whether not additionally accounting for secondary units (e.g., ads, search results, links) makes for misleading inference is more likely to depend on (partially unknown) deviations from an often implausible null hypothesis.  We illustrate that when experiments have effects, accounting for secondary units may be necessary to obtain trustworthy confidence intervals.

The literature on routine Internet-scale experimentation stresses the importance of running ``A/A tests'' as a validation of the combination of one's random assignment, data logging, and statistical inference procedures~\cite{crook2009seven,kohavi2012trustworthy,kohavi2009controlled}, though it is generally not stated exactly how these null experiments should be conducted and what their limitations are.  We intend that, in addition to our results, this paper provides a blueprint for other experimenters who wish to evaluate and choose among inferential procedures in their own settings.

\section{Dependence in experiments}
Many experiments allow observing the same units repeatedly: we may observe responses from the same person many times and also observe responses to the same items many times. In this section, we examine how this affects our estimation of contrasts between experimental conditions, such as differences in means between treatment and control.\footnote{
Online experiments may also exhibit other sources of dependence that are beyond the scope of this paper, including general equilibria in advertising auctions, peer effects, and other such ``spillovers.''}

Recent work in applied econometrics has been concerned with dependence due to clustering in data. It is now routine for work in empirical economics to consider and account for dependence in observations produced by one or more types of units.\footnote{For example, a recent paper by Cameron et al. \cite{cameron2006robust} on dealing with dependence due to observing two or more types of units has been cited over 600 times as of May 2013, according to Google Scholar.} Concerns about such dependence have been featured centrally in methodological work in the context of a growing number of field experiments in economics and other social sciences \cite{gerber2012field}. Similarly, work on two-way and tensor data in the context of recommender systems and observational comparisons has emphasized the importance of accounting for multiway dependency \cite{owen2007pigeonhole,owen2012bootstrapping}. And in psychometrics \cite{brennan1987bootstrap} and psycholinguistics \cite{baayen2008mixed}, investigators have identified problems with ignoring either of two sources of dependence.

As practitioners conducting and analyzing massive Internet experiments, the degree of attention given to this area suggests a need to consider the consequences of dependence for our data.  We present our efforts to understand whether it would be necessary to account for \emph{multiple} units causing dependence in our data, or whether a single unit would suffice in order to have inferential procedures with good performance.

\subsection{Generative models}
\label{sec:ranef_model}
In this section we describe a simple data generating process based on random effects models to illustrate how dependence can affect uncertainty in experiments and motivate the need to evaluate inferential procedures. 
Random effects models provide a general way to describe data arising from combinations of units, such as users and items (e.g., ads, search results).
In the two-way crossed random effects model~\cite{baayen2008mixed,searle1992variance}, each observation is generated by some function $f$ of a linear combination of a grand mean, $\mu$, a random effect $\alpha_i$ for the first unit, which (without loss of generality) we take to be the idiosyncratic deviation for user $i$, and a second random variable $\beta_j$ for the idiosyncratic deviation for item $j$.  Finally, we have a error term $\varepsilon_{ij}$ for each user's idiosyncratic response to each item.\footnote{For simplicity, we consider only a single observation of each user--item pair. Additional error terms can be included when there are repeated observations of pairs.}
This final term could be caused by a number of factors, including how relevant the item is to the user. Thus, we have the model
\begin{multline*}
Y_{ij} = f\left( \mu+ \alpha_i + \beta_j + \varepsilon_{ij}\right) \\
\alpha_i  \sim \mathcal{H}_{\alpha}(0, \sigma_{\alpha, i}^2), \hspace{.2cm} 
\beta_j \sim \mathcal{H}_{\beta}(0, \sigma_{\beta, j}^2), \hspace{.2cm} 
\varepsilon_{ij} \sim \mathcal{H}_{\varepsilon}(0, \sigma_{\varepsilon, {ij}}^2).
\end{multline*}
Each random effect is modeled as being drawn from some distribution with zero mean and some variance.  In the homogeneous random effects model, this variance is the same for each user or item (i.e., $\sigma_{\alpha, i} = \sigma_{\alpha}$), whereas in a heterogenous random effects model, each unit or group of units may have their own variances.

\subsubsection{Potential exposures and potential outcomes}
We generally do not observe all combinations of users and items; in fact, usually we only observe a small fraction of the possible combinations.  Which items users are exposed to may depend on user and item characteristics, and this pattern of exposure is often subject to experimental manipulation.  Without loss of generality, let users (rather than items) be randomly assigned to experimental conditions, so that $D_i$ is $i$'s assignment to a condition (e.g., in the case of a binary treatment, $D_i = 0$ is the control and $D_i = 1$ is the treatment). Let $Z^{(d)}$ be a matrix of indicator variables where $Z_{ij}^{(d)} = 1$ if and only if $i$ is exposed to item $j$ when $i$ is assigned to condition $d$.  The pattern of exposure $Z$ defines what outcomes can occur: if  $Z_{ij}^{(d)} = 1$ for some user--item-treatment combination $(i, j, d)$, then we would observe $Y_{i j}^{(d)}$ --- $i$'s \emph{potential outcome} in response to $j$ under the treatment $d$.  Note that since a user is only assigned to one condition, $D_i$, we cannot simultaneously observe both $Z_{i j}^{(0)}$ and $Z_{i j}^{(1)}$, nor can we observe both  $Y_{i j}^{(0)}$ and $Y_{i j}^{(1)}$.

\subsubsection{Difference in means for user--item experiments}
We wish to estimate quantities comparing outcomes that would occur under different values of $D_i$ --- most simply, the \emph{difference in means} for a binary treatment
\begin{align*}
\delta \equiv \hspace{.2cm} & \mathbb{E}[Y_{ij}^{(1)} \vert D_{i} = 1, Z_{ij}^{(1)} = 1] \; - \\  \hspace{.2cm} &  \mathbb{E}[Y_{ij}^{(0)} \vert D_{i} = 0, Z_{ij}^{(0)} = 1].
\end{align*}

Experiments can produce a non-zero $\delta$ simply by changing the pattern of users' exposure to items. For example, a search ranking experiment could primarily have effects by changing which items are displayed as results (and thus observed). At one extreme, it could be that the potential outcomes are identical under treatment and control, $Y_{i j}^{(0)} = Y_{i j}^{(1)}$ for all $i, j$, but that the pattern of exposure is different ($Z_{i j}^{(0)} \neq Z_{i j}^{(1)}$), such that $\delta \neq 0$.

Other experiments can produce a non-zero $\delta$ while leaving the pattern of exposure identical or otherwise ignorably similar. For example, an experiment might not alter which items are displayed to particular users, but instead render items slightly differently, so that $Y_{i j}^{(0)} \neq Y_{i j}^{(1)}$ for some $i, j$. In this case, $\delta$ is then an average treatment effect (ATE) since it is a difference in means for the same units \cite{rubin1974estimating}.

If for all $i, j$, $Z_{ij}^{(0)} =Z_{ij}^{(1)}$, the pattern of exposure is the same and $\delta$ is an ATE,
$$\delta = \mathbb{E}[Y_{ij}^{(1)} - Y_{ij}^{(0)} \vert Z_{ij} = 1].$$

For expository simplicity,\footnote{The results for the more general case require introducing a model for  exposure such that the random effects share common causes with the missingness. The $\sigma$ terms in variance expressions are then replaced with variances conditional on data being observed.} the remainder of this section assumes that the pattern of  exposure is the same under the treatment and control: $Z_{i j} \equiv Z_{i j}^{(0)} = Z_{i j}^{(1)}$.

\subsubsection{Estimates of average treatment effects}
We extend the basic random effects model above to include experimental conditions that may affect a user's exposure and response to an item.   We then derive expressions for the variance of the difference in mean outcomes between conditions to illustrate how repeatedly observing the same units, and which units are randomly assigned to conditions, influences estimates of experimental effects.

Here we restrict our attention to linear models with normally distributed random effects. That is, the following analysis considers cases where $Y$ is unbounded, $f$ is the identity function, and random effects are drawn from a multivariate normal distribution, so that
\begin{multline}
\label{random_effects_model_w_d}
Y_{ij}^{(d)} = \mu^{(d)} + \alpha_i^{(d)} + \beta_j^{(d)} + \varepsilon_{i j}^{(d)} \\
\vec{\alpha}_i  \sim \mathcal{N}(0, \Sigma_\alpha), \hspace{.2cm} 
\vec{\beta}_j \sim \mathcal{N}(0, \Sigma_\beta),  \hspace{.2cm} 
\vec{\varepsilon}_{ij} \sim \mathcal{N}(0, \Sigma_{\varepsilon}).
\end{multline}
Note that we define the random effects $\vec{\alpha}_i$, etc., as vectors with a covariance matrix (e.g., $\Sigma_{\alpha}$) so that their effects may be correlated across conditions.





The sample mean for each condition is
\begin{align*}
\begin{split}
\bar{Y}^{(d)} =\ & \mu^{(d)} +
\\ &\frac{1}{n_{\sumdot \sumdot}^{(d)}} \bigg(
 \sum_{i} n_{i \sumdot}^{(d)} \alpha_i^{(d)} + 
 \sum_{j} n_{\sumdot j}^{(d)} \beta_j^{(d)}   + 
 \sum_{i} \sum_{j} n_{i j}^{(d)} \varepsilon_{ij}^{(d)}
 \bigg)
 \end{split}
\end{align*}
where, e.g., $n_{j \sumdot}^{(d)}$ is the number of observations of item $j$ in condition $d$. We then estimate the ATE using the difference in observed sample means $\hat{\delta} = \bar{Y}^{(1)} - \bar{Y}^{(0)}$.

Consider the case where the number of observations in the treatment and control groups are of equal size such that $N = n_{\sumdot \sumdot}^{(1)} = n_{\sumdot \sumdot}^{(0)}$. This enables simplifying the expression for $\hat{\delta}$ and its variance to
\begin{align*}
\begin{split}
\hat{\delta} = \delta + \frac{1}{N} \bigg[
 &\sum_{i} \left( n_{i \sumdot}^{(1)} \alpha_i^{(1)} - n_{i \sumdot}^{(0)} \alpha_i^{(0)} \right) + \\
 & \sum_{j} \left( n_{\sumdot j}^{(1)} \beta_j^{(1)} - n_{\sumdot j}^{(0)} \beta_j^{(0)} \right) + \\
 & \sum_i \sum_j n_{ij}^{(D_{i})} \varepsilon_{ij}^{(D_{i})} \bigg]
\end{split}
\end{align*}
and
\begin{align}
\label{var_delta_hat_w_sums}
\begin{split}
\V[\hat{\delta}] =& \frac{1}{N^2} \bigg[
   \sum_i \bigg(
    \big(n_{i \sumdot}^{(1)}\big)^2  \sigma_{\alpha^{(1)}}^2 + 
      \big(n_{i \sumdot}^{(0)}\big)^2  \sigma_{\alpha^{(0)}}^2 \bigg) \\
  + & \sum_j \bigg(
    \big(n_{\sumdot j}^{(1)}\big)^2  \sigma_{\beta^{(1)}}^2 + 
      \big(n_{\sumdot j}^{(0)}\big)^2  \sigma_{\beta^{(0)}}^2 -
    2 n_{\sumdot j}^{(0)} n_{\sumdot j}^{(1)} \sigma_{\beta^{(0)}, \beta^{(1)}}^2 \bigg) \\
  + & \sum_i \sum_j \bigg(\big(n_{ij}^{(1)}\big)^2 \sigma_{\varepsilon^{(1)}} - \big(n_{ij}^{(0)}\big)^2 \sigma_{\varepsilon^{(0)}}  \bigg)
  \bigg].
  \end{split}
\end{align}
The first term in (\ref{var_delta_hat_w_sums}) is the contribution of random effects of users to the variance, and the second is the contribution of the random effects of items. The covariance term $\sigma_{\beta^{(0)}, \beta^{(1)}}^2$, present for items, is absent for users and user--item pairs since each is only observed in either the treatment or control.

To further simplify, we can introduce coefficients measuring how much units are duplicated in the data. Following previous work \cite{owen2007pigeonhole,owen2012bootstrapping}, we define
\begin{align*}
&\nu_A^{(d)} \equiv \frac{1}{N} \sum_i \big(n_{i \sumdot}^{(d)} \big)^2
&\nu_B^{(d)} \equiv \frac{1}{N} \sum_j \big(n_{\sumdot j}^{(d)} \big)^2,
\end{align*}
which are the average number of observations sharing the same user (the $\nu_A$s) or item (the $\nu_B$s) as an observation (including itself). For the units assigned to conditions (in this case, users), either $n_{i \sumdot}^{(0)}$ or $n_{i \sumdot}^{(1)}$ is zero for each $i$; for the non-assigned units (items), we need a measure of this between-condition duplication
\begin{align*}
&\omega_B \equiv \frac{1}{N} \sum_j n_{\sumdot j}^{(0)} n_{\sumdot j}^{(1)}.
\end{align*}
Under the homogeneous random effects model (\ref{random_effects_model_w_d}), we can then simplify (\ref{var_delta_hat_w_sums}) to
\begin{align}
\begin{split}
\V[\hat{\delta}] =& \frac{1}{N} \bigg[
   \bigg( \nu_A^{(1)} \sigma_{\alpha^{(1)}}^2  + \nu_A^{(0)} \sigma_{\alpha^{(0)}}^2 \bigg) \\
  &+ \bigg( \nu_B^{(1)} \sigma_{\beta^{(1)}}^2 + \nu_B^{(0)} \sigma_{\beta^{(0)}}^2 
  - 2 \omega_{\beta} \sigma_{\beta^{(0)}, \beta^{(1)}}^2 \bigg) \\
  &+ \bigg( \sigma_{\varepsilon^{(0)}}^2 + \sigma_{\varepsilon^{(1)}}^2 \bigg)
  \bigg].
\end{split}
\end{align}
The above expression makes clear that if the random effects for items in the treatment and control are correlated (as we would usually expect), then an increase in the balance of how often items appear in each condition reduces the variance of the estimated treatment effect.

\subsubsection{Sharp and non-sharp null hypotheses}
{\bf Sharp null hypothesis}. Under the \emph{sharp null hypothesis} for user--item experiments, the treatment has no average, interaction, or exposure effects; that is, the outcome for a particular user--item pair, and whether or not the item is displayed to the user, would be the same regardless of treatment assignment. 
In the context of our model, in addition to $\delta = 0$, the sharp null can be defined by:
\begin{align}
\label{sharp_null_layout}
\begin{split}
& Z_{i j} \equiv Z_{i j}^{(0)}= Z_{i j}^{(1)}
\end{split}
\end{align}
\begin{align}
\label{sharp_null_variances}
\begin{split}
&\sigma_{\alpha}^2 \equiv \sigma_{\alpha^{(1)}}^2 = \sigma_{\alpha^{(0)}}^2 = \sigma_{\alpha^{(0)}, \alpha^{(1)}}^2 \\ 
&\sigma_{\beta}^2 \equiv \sigma_{\beta^{(1)}}^2 = \sigma_{\beta^{(0)}}^2 = \sigma_{\beta^{(0)}, \beta^{(1)}}^2 \\
&\sigma_{\varepsilon}^2 \equiv \sigma_{\varepsilon^{(1)}}^2 = \sigma_{\varepsilon^{(0)}}^2 = \sigma_{\varepsilon^{(0)}, \varepsilon^{(1)}}^2.
\end{split}
\end{align}

In the case of the sharp null, only random effects for items that are not balanced across conditions contribute to the variance of our difference: the contribution a single item $j$ makes to the variance simplifies to $\big(n_{\sumdot j}^{(0)} - n_{\sumdot j}^{(1)} \big)^2 \sigma_{\beta}^2$; that is, it depends only on the squared difference in duplication between treatment and control. It is easy to show that
\begin{align}
\label{ranef_var_under_sharp_null}
\begin{split}
\V[\hat{\delta}] =& \frac{1}{N} \bigg[
   \big( \nu_A^{(1)} + \nu_A^{(0)} \big) \sigma_{\alpha}^2
  + \kappa_B \sigma_\beta^2 + 2 \sigma_{\varepsilon}^2
  \bigg],
\end{split}
\end{align}
where $\kappa_B \equiv \frac{1}{N} \sum_j \big(n_{\sumdot j}^{(1)} - n_{\sumdot j}^{(0)} \big)^2$ measures the average between-condition duplication of observations of items. If items, like users, also only appear in either treatment or control, then $\kappa_B = \nu_B^{(1)} + \nu_B^{(0)}$, highlighting the resulting symmetry between users' and items' contributions to our uncertainty.

{\bf Non-sharp null hypothesis}.
Experiments may have zero average effect ($\delta = 0$) and still violate the sharp null.  For example, when (\ref{sharp_null_layout}) does not hold, the pattern of exposure may change such that users are exposed to different items in each condition, but there are no user or item effects (i.e., (\ref{sharp_null_variances}) remains true).  This can occur when exposures are missing from the layout $Z^{(d)}$ at random. That is, the change in exposure between conditions is independent of the potential outcomes: $Z^{(1)}_{i j} - Z^{(0)}_{i j} \independent (Y^{(0)}_{i j}, Y^{(1)}_{i j})$. 

Another deviation from the sharp null is when (\ref{sharp_null_variances}) does not hold.  In this case, we say that there are interaction effects of the treatment and units; for example, an experiment can positively or negatively affect particular items or user--item pairs, but when outcomes are averaged over all exposures, there is zero effect.

Together these considerations highlight the ways in which field experiments without true average effects can appear to have differences between treatment and control conditions due to effects of users and items, imbalance, and treatment-item interactions.  Inferential methods that do not account for these kinds of dependence structures may result in understating the variance of the difference of means estimator used to produce confidence intervals.  In the next section we review bootstrap methods which can account for varying levels of dependence structure.




\subsection{Bootstrapping dependent data}\label{sec:bootstrapping}
The bootstrap \cite{efron1979bootstrap} offers a very general method for characterizing the sampling distribution of a statistic (e.g., a difference in means), and can be used to produce confidence intervals for experimental comparisons for many data generating processes.
The bootstrap distribution of a sample statistic is the distribution of that statistic under resampling \cite{efron1979bootstrap} or reweighting \cite{rubin1981bayesian} of the data. In this section, we describe how the bootstrap can be applied to dependent data. We focus on a version of the bootstrap that uses independent weights, rather than the resampling bootstrap,
since it is suitable for use in online (i.e., streaming) computational settings \cite{owen2012bootstrapping,oza2001experimental}.

Bootstrap methods are attractive because they involve minimal assumptions and scale well to large datasets compared to other methods commonly used in practice for statistical inference with dependent data. One could fit crossed random effects model to the data \cite{baayen2008mixed}, but such models don't scale to large datasets and involve specifying (likely incorrect) assumptions not needed for the bootstrap.   Other alternatives include cluster robust Huber--White ``sandwich'' standard errors from econometrics
\cite{cameron2006robust}, but such methods cannot be applied in a streaming fashion and are not available for robust statistics of interest, such as trimmed or Winsorized means.



\subsubsection{iid bootstrap}
In order to get a confidence interval for some statistic $t$, we produce $R$ replicates of the the statistic, $t_{r}^{*}$,
computed on randomly reweighted versions of the sample. 
That is, for some replicate $r \in [1, R]$, each observation $Y_{ij}$ is randomly reweighted with weights $W_{ijr}$. These reweighted samples allow us to estimate features of the sampling distribution of our statistic. We generally have
$W_{ijr} \sim \mathcal{G}$
where $\mathcal{G}$ is some distribution with mean and variance 1, such as $\text{Poisson}(1)$ and $\text{Uniform}\{0, 2\}$ \cite[\S 3.3]{owen2012bootstrapping}.
Note that each observation is reweighted independently, including other observations of the same units. Applied to two-way data, the iid bootstrap can be expected to underestimate the variance of statistics and thus produce confidence intervals with poor coverage~\cite{mccullagh2000resampling}.

\begin{figure*}[ht!]
\begin{center}
\includegraphics[width=.94\textwidth]{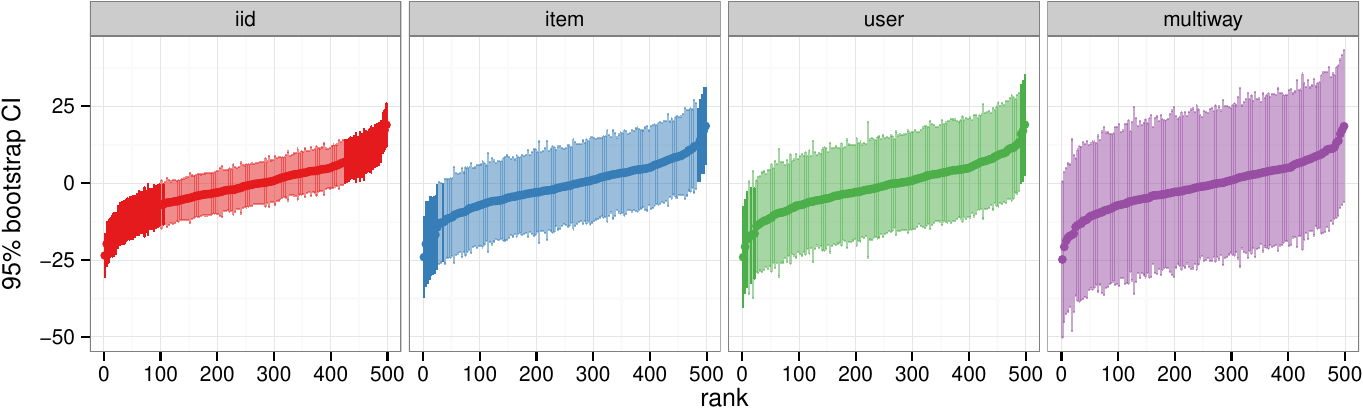}
\caption{An illustration of our method for computing true coverage rates for the bootstrap methods with the Search dataset. We compute null experiments to obtain nominal ``95\% confidence intervals'' for the difference in means $\hat{\delta}_{kr}$, and count the fraction of tests that accept the null hypothesis (e.g., indicate there is no significant difference in means). To show how results can vary between comparisons, we sort the results by $\mathbb{E}_{r}[\hat{\delta}_{kr}]$, and darken results that (incorrectly) reject the null. Anti-conservative tests -- in this case, the iid and item-clustered bootstrap -- reject in more than 5\% of the experiments.  Differences in the figure are shown relative to the grand mean.}
\vspace{-11pt}
\label{fig:ci_width}
\end{center}
\end{figure*}

\subsubsection{One-way bootstrap}
In the one-way bootstrap, or ``block'' or ``cluster'' bootstrap, the analyst chooses a single relevant type of unit (e.g., users) and all observations from the same unit are given the same random weight when reweighting.
In other words, taken $i$ as indexing the chosen type of unit, we have $W_{ijr} = W_{ij'r} = u_{ir}$ and $u_{ir} \sim \mathcal{G}$ for all $j, j'$. When the data only has one-way dependency, this procedure produces a bootstrap distribution that gives consistent confidence intervals. When the data has additional dependency structure, it can be anti-conservative.

\subsubsection{Multiway bootstrap}
When there are two or more relevant units, analysts can use a bootstrap that reweights all relevant units. Under a more general model than the one presented above, the multiway bootstrap produces variance estimates, and thus confidence intervals, that are accurate or mildly conservative \cite{owen2007pigeonhole,owen2012bootstrapping}.
The two-way bootstrap has been used for analyzing large online advertising experiments~\cite{bakshy2012social}.   With two-way data, we have $W_{ijr} = u_{ir} v_{jr}$, where $u_{ir} \sim \mathcal{G}$ and $v_{jr} \sim \mathcal{G}$. That is, the random weights for an observation is the product of two independently sampled weights assigned to unit $i$ and unit $j$. For example, if in one replicate, user $i$ gets weight $2$ and item $j$ gets weight $3$ then all observations of the pair $(i, j )$ get weight $2 \times 3 = 6$ in that replicate.   Note that if either unit has a weight of 0, any combination of that unit with another unit will be given weight of 0.  This procedure can be generalized to cover $d$-way data in a straightforward fashion~\cite{owen2012bootstrapping}.

\subsubsection{Online bootstrap computation}
For any statistic $t$ that can be computed online, the one-way bootstrap can be implemented online as follows \cite{owen2012bootstrapping,oza2001experimental}: on visiting each observation, use a hash of an identifier of each unit (e.g., a user ID) as the seed to a pseudorandom number generator for $\mathcal{G}$, draw $R$ weights, one for each bootstrap replicate $r$, and use these weights to update the running sufficient statistics for $t^*_r$.  The $d$ dimensional multiway bootstrap can be implemented using a similar procedure using the products of weights generated from each unit.

\section{Empirical Evaluation}
Quantifying the uncertainty in experimental effects requires that we correctly estimate the variance of a difference in means.  Based on our intuitions from the simple model in Section~\ref{sec:ranef_model}, this variance can depend critically on at least four sources of variation: effects of users and items, their duplication, how well items are balanced across conditions, and heterogeneity in treatment effects.  We might expect that inferential procedures which do not account for users and items to produce inaccurate confidence intervals; we explore this hypothesis with respect to these four sources in turn in using synthetic null experiments in the remaining sections.

\subsection{Data}
\label{sec:description}
We examine click-through rate outcomes for three core product areas at Facebook: Ads, Search, and News Feed. 

{\bf Ads}. We analyze ad click-through rates for one type of ad unit for a popular advertising product on Facebook. Each impression corresponds to a single delivery of the ad to a user's Web browser.

{\bf Search}. We analyze search click-through rates for one type of search result on Facebook.  Each impression is a validated delivery of an item in the ``typeahead'' results, and each click is a click on the item.  Note that if an item presented multiple times over several query reformulations, each is considered a separate impression.

{\bf Feed}.  We analyze click-through rates for one type of story in the News Feed in a large country.  Each impression corresponds to a single delivery of the story to a viewer's Web browser, and a click corresponds to a click on the item's thumbnail or snippet.

\subsection{Basic A/A test}\label{sec:computation}
The most basic A/A test we consider is a synthetic experiment that evaluates inferential methods under the sharp null by partitioning the data into multiple random segments and computing confidence intervals for the difference in mean outcomes several hundred times.  To do this, we first take the identifier of the unit we wish to randomize over (the user id), concatenate it with a \emph{salt} (i.e., an integer), compute this string's MD5 hash value, and assign the unit to a segment number that is the integer representation of the first 7 digits of the hash modulo $M=100$.  We use a similar procedure to hash the secondary units (items). We generate the confidence intervals for differences between even and odd numbered segments, yielding 50 comparisons per salt, and repeat this procedure for 10 salts, yielding $K = 500$ null experiment comparisons, illustrated in Figure~\ref{fig:ci_width}.

\begin{figure*}[!th]
\begin{center}
\includegraphics[width=0.95\textwidth]{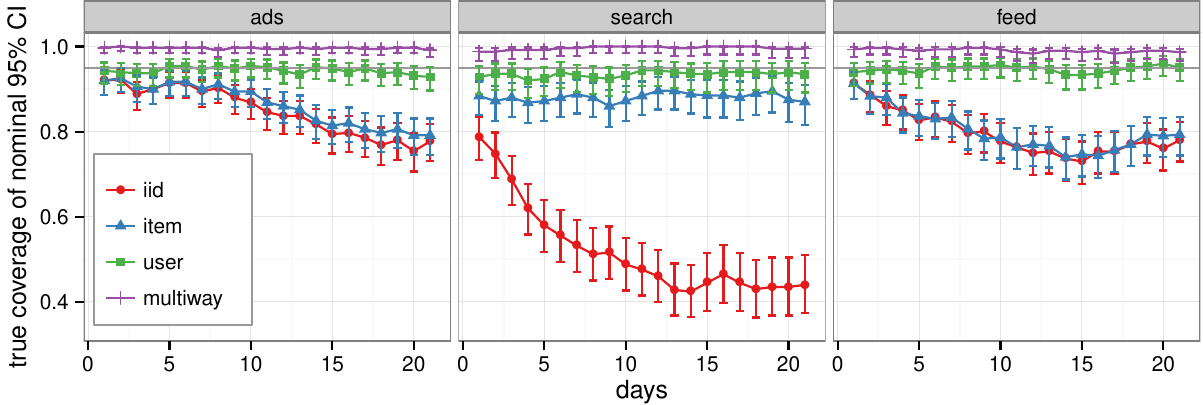}
\caption{True coverage for nominal 95\% confidence intervals produced by the iid, one-way, and multiway bootstrap for A/A tests segmented by user id as a function of time.  Uncertainty estimates for the iid and item-level bootstrap become increasingly inaccurate over time, while the user-level and multiway bootstrap have the advertised or conservative Type I error rate.}
\vspace{-10pt}
\label{fig:coverage_time}
\end{center}
\end{figure*}

The confidence intervals for each bootstrap method for each null experiment comes from $R=500$ bootstrap reweightings of the data. To determine whether or not an experiment is significant, we compute the mean and variance of the difference in means $\hat{\delta}_{kr}$ over all $R$ replicates for each of the $K$ experiments. The distributions of $\hat{\delta}_{kr}$ are asymptotically normal under the bootstrap, so we simply use normal quantile function to compute the central $100(1 - \alpha)\%$ interval.

To obtain the estimated \emph{true coverage} for the nominal 95\% confidence intervals, we compute the proportion of times the $K$ bootstrap tests indicate a significant difference in means at $\alpha = 0.05$. We treat each of the $K$ comparisons as independent, and use the Wilson score interval for binomial proportions~\cite{agresti2002categorical} to determine the uncertainty around the estimated true coverage rate.

\subsection{Duplication}
\label{sec:duplication}

A central quantity that contributes to the variance of  $\hat{\delta}$ is the average number of observations that share the same user, $\nu_{\text{user}}$, and item, $\nu_{\text{item}}$.  We give basic summary statistics about the duplication in the data for a random 1\% segment used in our evaluation for the three datasets in Table~\ref{tab:summary}.  For the restricted categories of items we consider in each dataset, there are more users exposed to ads than the search results or feed stories.  While per-user duplication is similar across the three datasets, the per-item duplication for Ads is much higher than either Search or Feed.  This pattern is congruent with the nature of the items: the number of businesses that are actively advertising are far fewer than the number of users, while search and News Feed stories tend to have a much longer tail of items that result in lower duplication.

\begin{table}[t]

\begin{center}
\begin{tabular}{lrrr}
 \hline
 & Ads & Search & Feed  \\ 
   \hline \vspace{1pt}
  users & 4,515,816 &  908,339 &  545,218 \\ \vspace{1pt}
  items &  317,159 & 1,362,061 &  326,831 \\ \vspace{1pt}
  user--item pairs & 24,081,939 &  4,263,769 &  2,882,452 \\ \vspace{1pt}
  $\nu_{\text{users}}$ & 18.5 & 35.5 & 20.3 \\ \vspace{1pt}
  $\nu_{\text{items}}$ & 6,625.9 &  543.6 & 1,333.0 \\
   \hline
\end{tabular}
\end{center}
\caption{The amount of duplication present in our datasets for a single 1\% segment of users.}
\label{tab:summary}
\end{table}

\begin{figure}[t]
\includegraphics[width=.99\columnwidth]{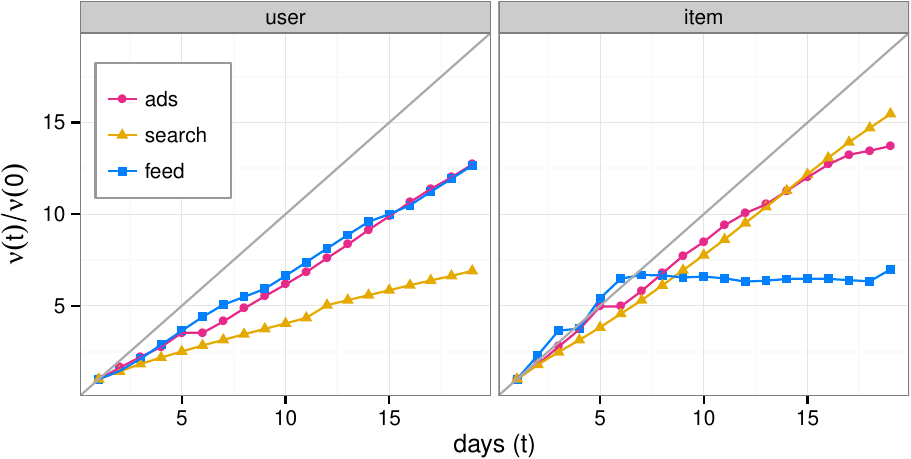}
\caption{Duplication ($\nu$) for users and items over time relative to the first day.}
\label{fig:duplication_time}
\end{figure}

Experiments often run for many days; as the number of days increase, so does the duplication.  Figure~\ref{fig:duplication_time} shows how duplication increases over time.  With the exception of Feed items, this relationship is rather linear both for user and items.  The behavior for Feed may be explained by the way social media feeds work: unlike ads and search results, users see and interact with very recent content, therefore limiting the average number of users that may be exposed to an item.

The sharp null variance expression (\ref{ranef_var_under_sharp_null}) suggests that we might expect that the increase in user and item-level duplication over time to contribute substantially to the variance of $\hat{\delta}$.  Not taking these units into account when computing confidence intervals may result in poor coverage.  Figure~\ref{fig:coverage_time} shows the true coverage of the different bootstrap methods for consecutively larger spans of time in each dataset.  We find that the iid confidence intervals tend to be highly anti-conservative.  For example, after two weeks of data collection, a search experiment that tests the difference in click-through rates between two equivalent groups of users could result in rejecting the null hypothesis nearly 50\% of the time.  We find that bootstrapping by the unit not being randomized over (the item) often leads to anti-conservative intervals, and that for the sharp null, which has little item imbalance or item-treatment effects, the user-level bootstrap yields accurate coverage.  The multiway bootstrap, 
however,
remains conservative no matter how many days are considered.

\subsection{Imbalance in items}
\label{sec:imbalance}
Given how the synthetic experiments in Section~\ref{sec:computation} were constructed, there is approximate balance of items across conditions such that the primary contributors to the variance of {$\hat{\delta}$ are expected to come from user and residual error random effects.  However, if items are systematically imbalanced across treatments (e.g., the experiment results in showing similarly relevant, but different ads), then based on our intuition from (\ref{ranef_var_under_sharp_null}), item random effects can also make a substantial contribution.

To examine how such imbalance might affect the coverage of the confidence intervals, we created imbalance by downsampling items from either condition with probability $p$.  This type of data augmentation results in synthetic experiments that correspond to the non-sharp null hypothesis with zero mean difference, equal variances, and \emph{different} patterns of potential exposure ($Z^{(0)} \neq Z^{(1)}$).

To do this, for each pair of segments $(m, m+1)$, we downsample each item from either segment even or odd segments (chosen with equal probability); in the downsampled segment for some item $j$, its user--item pairs are independently removed with probability $p$.
Thus, when $p=0$, we have the sharp null hypothesis, and when $p=1$, we have total imbalance (i.e., the two conditions contain disjoint sets of items).

\begin{figure}[t]
\begin{center}
\vspace{.25\columnwidth}
Figure withdrawn due to error.
\vspace{.25\columnwidth}
\caption{True coverage for nominal 95\% confidence intervals for each bootstrap method applied to data with varying levels of synthetic imbalance of items across conditions for 2 weeks of data.
This figure has been withdrawn because of an error in the software used to create synthetic imbalance of units.
}
\end{center}
\label{fig:censor_data}
\end{figure}

Figure~4 previously reported the true coverage with varying censoring probabilities, $p \in \{0.3, 0.6, 1.0\}$.
\emph{However, we have identified an error in the software used to produce this synthetic imbalance, making these results in error.} Thus, we have removed the text that described these results from this updated version of this paper.



\section{Simulations with effects}
\label{sec:simulation}

\begin{figure*}
\begin{center}
\includegraphics[width=.81\textwidth]{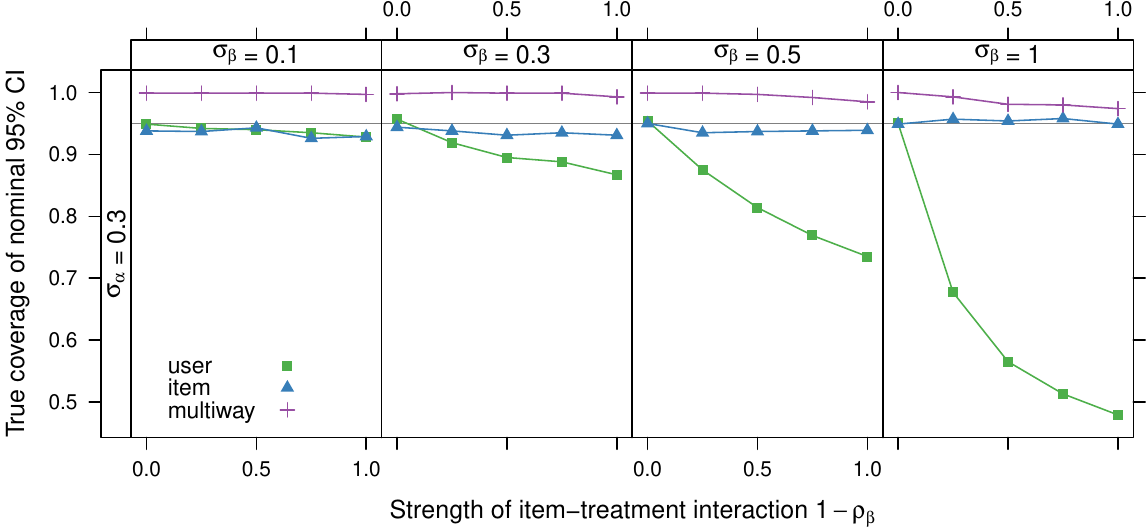}
\caption{Effects of item--treatment interaction effects on true coverage for nominal 95\% confidence intervals. Decreasing $\rho_\beta$, which makes the random item effects less correlated between treatment and control, reduces the coverage of user bootstrap confidence intervals. This effect is moderated by the magnitude of the item-level random effects.}
\label{fig:sim_interaction_coverage}
\end{center}
\end{figure*}
We have seen how different bootstrap methods perform in practice under the sharp null where experiments have no effects at all. However, these two tests cannot tell us about how bootstrap procedures might perform in situations where experiments affect potential outcomes for specific user--item pairs.  For example, an ads experiment that manipulates the display of certain advertising units may only affect certain items and not others~\cite{bakshy2012social}.  To explore these circumstances, we conduct simulations with a probit random effects model parameterized to mirror the kinds of outcomes described in the previous section. We use this generative model to vary the presence of an item--treatment interaction, a plausible source of violations of the sharp null hypothesis given in (\ref{sharp_null_variances}).

We modify the model of (\ref{random_effects_model_w_d}) so that $Y$ is binary and there is a single intercept common to both treatment and control, reflecting the lack of an ATE:
\begin{align}
\nonumber y_{ij}^{(d)} &= \mu + \alpha_i^{(d)} + \beta_j^{(d)} + \varepsilon_{i j}^{(d)} \\
\nonumber \mathbb{E}[Y_{ij}^{(d)}] &= \mathbbm{1}{\{y_{ij}^{(d)} > 0\}}
\end{align}
Also reflecting the absence of an ATE, we restrict the random effect variance to be the same in treatment and control. For example, the covariance matrix for the item random effects is
$$
\Sigma_{\beta} = 
\left[ \begin{array}{c c}
    \sigma_\beta^2 &  \rho_\beta \sigma_\beta^2 \vspace{.2cm} \\ 
    \rho_\beta \sigma_\beta^2 &  \sigma_\beta^2
\end{array} \right].
$$
To make realistic choices for the variances of the random effects, we fit a probit random effects models to the ads dataset from a large random sample of users in each of several small countries. This produced several estimates of $\sigma_\alpha$ and $\sigma_\beta$. We report on simulation results for $\sigma_\alpha = 0.3$, which is close to several of the estimates. Our estimates of $\sigma_\beta$ often ranged from 0.2 to 0.9, so we present results for $\sigma_\beta \in \{0.1, 0.3, 0.5, 1.0\}$.  We set $\mu$ so as to achieve $\mathbb{E}[Y_{ij}]$ close to $0.02$.\footnote{Since there is no scale to the latent variable $y_{ij}$, we achieved this by in fact choosing a fixed $\mu = -2$ and rescaling the random effect variances to sum to 1.}

We constructed the set of observed user--item pairs used in the simulations by assigning each of 3,000 potential users and 200 potential ads to log-normally distributed scores. For each of $2N$ observations, we selected a particular user and ad with probability proportional to this score. This yielded a ``layout'' with 2481 unique users, 199 unique ads, and duplication coefficients $\nu_A \doteq 30.9$ and $\nu_B \doteq 6077.4$, which is similar to the Ads dataset.




\subsection{Item--treatment interactions}
Even if the treatment has no effects on average, it can have positive effects for some users and items and negative effects for others. Given our random effects model, we know that item--treatment interactions can increase the contribution of duplication of items to the variance of the mean difference.

We vary item--treatment interactions by setting the correlation coefficient $\rho_\beta \in \{0, 0.25, 0.5, 0.75, 1\}$. Perfect correlation $\rho_\beta = 1$ corresponds to data generated under the sharp null hypothesis, while decreasing $\rho_\beta$ corresponds to an increasing proportion of item random effects being not shared across conditions. At the extreme of $\rho_\beta = 0$, the random effect of an item in the treatment is completely independent of its random effect in the control.

\subsection{Results}
Figure \ref{fig:sim_interaction_coverage} summarizes the results of 1000 simulations for each combination of parameter values. Without any item--treatment interaction, both the user and item bootstrap have approximately correct coverage; this is attributable to the relatively low $\nu_A$ in this simulation, and is consistent with the results from a small number of days in our datasets. As the item--treatment interaction increases, the coverage of the user bootstrap confidence intervals drops substantially. For example, even with moderate values of $\sigma_\beta = 0.5$ and $\rho_\beta = 0.75$, a nominally 95\% confidence interval has a true coverage of 87.5\%. While do we not expect to observe the extremes of all item-level variance being treatment specific (i.e., $\rho_\beta = 0$), these results demonstrate that deviations from the sharp null in the form of item--treatment interaction have serious consequences for the one-way bootstrap. On the other hand, the multiway bootstrap remains mildly conservative even with large $\sigma_\beta$ and small $\rho_\beta$.

\section{Discussion}
Despite having a large number of individual observations, many settings for online experiments involve substantial dependence and small effects such that statistical inference remains a central concern. The preceding analysis of real and simulated data makes clear that methods which neglect dependence structure in these large experiments can result in high Type I error rates and confidence intervals with poor coverage. In each of our three datasets, the iid bootstrap performed very poorly, such that using it (or other methods assuming iid observations) would result in reaching incorrect conclusions about the presence, sign, and magnitude of treatment effects~\cite{gelman2000type}.   Furthermore, the particulars of the user--item exposure layout ($Z$) provide new considerations that deserve further attention.

On the other hand, neglecting dependence among observations of units not assigned to conditions (the items) generally did not result in lower coverage with our data.

Neglecting dependence among observations of non-experimental units (e.g., items) may have substantial effects on coverage when the treatment has any effects. Most treatments are expected to have some effects. Our simulations with item--treatment interaction effects demonstrate that the coverage of the user bootstrap can be extremely sensitive to the presence of these effects. This highlights that using A/A tests only serves to validate inferential
procedures under a narrow set of conditions (i.e., the sharp null hypothesis),
but cannot detect other (potentially severe) inferential problems that
occur in other circumstances. Given that experimenters expect treatment effects, and often want to know how large the average effects are, they should consider whether or not they wish to use a procedure that provides a somewhat conservative measurement of uncertainty (i.e., the multiway bootstrap), or the user-level bootstrap, which correctly tests the less plausible sharp null.

A limitation of the present work is that, from the perspective of experimenters such as ourselves trying to evaluate inferential methods in practice, there is remaining gap between what is possible to learn from straightforward perturbations of real datasets and what is possible to learn from necessarily simplified generative models. Future work may develop more sophisticated ways of perturbing existing data and using additional parameters estimated from real experiments to produce evaluations for data that more closely resemble outcomes in the field.

This paper has been primarily concerned with Type I error rates and the coverage of confidence intervals, but experimenters are equally concerned about Type II errors (failures to reject the null) and related errors such as incorrectly estimating the direction or magnitude of effects. Many principled approaches to
choosing how to assign units to one of many available treatments over time (e.g., solutions to multi-armed bandit problems) require correctly estimating one's uncertainty about the expected payoffs of the treatments~\cite{scott2010modern}. Therefore, we expect that addressing multiway dependence will remain important when taking these approaches as well.
A related point is that experimenters often exert considerable effort \emph{reducing} the width of CIs by increasing precision through design and adjustment \cite{box2005statistics,lin_agnostic_2013,miratrix2012adjusting}. Many of these methods could be applied in combination with single or multiway bootstrapping. Finally, there may be other practical ways to reduce the width of multiway bootstrap CIs through using linear combinations of variance estimates from different bootstrap procedures \cite{cameron2006robust,owen2012bootstrapping}.

\section{Acknowledgements}
We would like to give many thanks to Daniel Merl, as well as Alex Deng, Wojciech Galuba, Brian Karrer, Art B. Owen, Luca Pozzi, and Daniel Ting for their thoughtful comments.

\bibliographystyle{abbrv}
\bibliography{boot_eval} 

\begin{thebibliography}{10}

\bibitem{agresti2002categorical}
A.~Agresti.
\newblock {\em Categorical Data Analysis}.
\newblock {Wiley-Interscience}, 2nd edition, July 2002.

\bibitem{baayen2008mixed}
R.~H. Baayen, D.~J. Davidson, and D.~M. Bates.
\newblock Mixed-effects modeling with crossed random effects for subjects and
  items.
\newblock {\em Journal of Memory and Language}, 59(4):390--412, 2008.

\bibitem{bakshy2012social}
E.~Bakshy, D.~Eckles, R.~Yan, and I.~Rosenn.
\newblock Social influence in social advertising: evidence from field
  experiments.
\newblock In {\em Proceedings of the 13th ACM Conference on Electronic
  Commerce}, pages 146--161. ACM, 2012.

\bibitem{box2005statistics}
G.~E. Box, J.~S. Hunter, and W.~G. Hunter.
\newblock {\em Statistics for Experimenters: Design, Innovation, and
  Discovery}, volume~13.
\newblock Wiley Online Library, 2005.

\bibitem{brennan1987bootstrap}
R.~L. Brennan, D.~J. Harris, and B.~A. Hanson.
\newblock {\em The bootstrap and other procedures for examining the variability
  of estimated variance components in testing contexts}.
\newblock American College Testing Program, 1987.

\bibitem{cameron2006robust}
A.~Cameron, J.~Gelbach, and D.~Miller.
\newblock Robust inference with multi-way clustering.
\newblock {\em Journal of Business \& Economic Statistics}, 29(2):238--249,
  2011.

\bibitem{crook2009seven}
T.~Crook, B.~Frasca, R.~Kohavi, and R.~Longbotham.
\newblock Seven pitfalls to avoid when running controlled experiments on the
  web.
\newblock In {\em Proceedings of the 15th ACM SIGKDD international conference
  on Knowledge discovery and data mining}, pages 1105--1114. ACM, 2009.

\bibitem{efron1979bootstrap}
B.~Efron.
\newblock Bootstrap methods: Another look at the jackknife.
\newblock {\em The Annals of Statistics}, 7(1):1--26, 1979.

\bibitem{gelman2000type}
A.~Gelman and F.~Tuerlinckx.
\newblock Type {S} error rates for classical and {B}ayesian single and multiple
  comparison procedures.
\newblock {\em Computational Statistics}, 15(3):373--390, 2000.

\bibitem{gerber2012field}
A.~S. Gerber and D.~P. Green.
\newblock {\em Field Experiments: Design, Analysis, and Interpretation}.
\newblock WW Norton, 2012.

\bibitem{kohavi2012trustworthy}
R.~Kohavi, A.~Deng, B.~Frasca, R.~Longbotham, T.~Walker, and Y.~Xu.
\newblock Trustworthy online controlled experiments: Five puzzling outcomes
  explained.
\newblock In {\em Proceedings of the 18th ACM SIGKDD international conference
  on Knowledge discovery and data mining}, pages 786--794. ACM, 2012.

\bibitem{kohavi2009controlled}
R.~Kohavi, R.~Longbotham, D.~Sommerfield, and R.~Henne.
\newblock Controlled experiments on the web: Survey and practical guide.
\newblock {\em Data Mining and Knowledge Discovery}, 18(1):140--181, 2009.

\bibitem{lin_agnostic_2013}
W.~Lin.
\newblock Agnostic notes on regression adjustments to experimental data:
  Reexamining {F}reedman's critique.
\newblock {\em The Annals of Applied Statistics}, 7(1):295--318, 2013.

\bibitem{mccullagh2000resampling}
P.~McCullagh.
\newblock Resampling and exchangeable arrays.
\newblock {\em Bernoulli}, 6(2):285--301, 2000.

\bibitem{miratrix2012adjusting}
L.~W. Miratrix, J.~S. Sekhon, and B.~Yu.
\newblock Adjusting treatment effect estimates by post-stratification in
  randomized experiments.
\newblock {\em JR Stat. Soc. Ser. B. Stat. Methodol. To appear}, 2012.

\bibitem{owen2007pigeonhole}
A.~B. Owen.
\newblock The pigeonhole bootstrap.
\newblock {\em The Annals of Applied Statistics}, 1(2):386--411, 2007.

\bibitem{owen2012bootstrapping}
A.~B. Owen and D.~Eckles.
\newblock Bootstrapping data arrays of arbitrary order.
\newblock {\em The Annals of Applied Statistics}, 6(3):895--927, 2012.

\bibitem{oza2001experimental}
N.~C. Oza and S.~Russell.
\newblock Experimental comparisons of online and batch versions of bagging and
  boosting.
\newblock In {\em Proceedings of the seventh ACM SIGKDD international
  conference on Knowledge discovery and data mining}, pages 359--364. ACM,
  2001.

\bibitem{rubin1974estimating}
D.~B. Rubin.
\newblock Estimating causal effects of treatments in randomized and
  nonrandomized studies.
\newblock {\em Journal of Educational Psychology}, 66(5):688--701, 1974.

\bibitem{rubin1981bayesian}
D.~B. Rubin.
\newblock The {B}ayesian bootstrap.
\newblock {\em The Annals of Statistics}, 9(1):130--134, 1981.

\bibitem{scott2010modern}
S.~L. Scott.
\newblock A modern {B}ayesian look at the multi-armed bandit.
\newblock {\em Applied Stochastic Models in Business and Industry},
  26(6):639--658, 2010.

\bibitem{searle1992variance}
S.~R. Searle, G.~Casella, C.~E. McCulloch, et~al.
\newblock {\em Variance Components}.
\newblock Wiley New York, 1992.

\end{thebibliography}

\end{document}